\documentclass{optica-article}

\journal{opticajournal} 

\articletype{Research Article}

\usepackage{lineno}
\usepackage{float}

\begin{document}

\title{6-mJ, 4-ns Pulse Generation at 2.09~\textmu m from a Diode-Pumped Ho:YAG Thin-Disk Laser}

\author{Yuya Koshiba,\authormark{1,*} Jiří Mužík,\authormark{1} Martin Smrž,\authormark{1} Matyáš Dvořák,\authormark{2}, Sabina Kudělková,\authormark{3} Antonín Fajstavr,\authormark{3} and Tomáš Mocek\authormark{1}}

\address{\authormark{1}HiLASE Centre, Institute of Physics of the Czech Academy of Sciences, Za Radnicí 828, 252 41 Dolní Břežany, Czechia\\
\authormark{2}Celestia Energy s r.o., 5. května 16, 252 41 Dolní Břežany, Czechia\\
\authormark{3}CRYTUR, spol. s r.o., Na Lukách 2283, 511 01 Turnov, Czechia}

\email{\authormark{*}yuya.koshiba@hilase.cz} 


\begin{abstract*} 
A holmium-doped yttrium aluminum garnet (Ho:YAG) thin-disk was experimentally investigated under Q-switching and cavity-dumping operation schemes, pumped by a 1.9~\textmu m laser-diode (LD). The laser generated pulses at 2090~nm with energies more than 6~mJ and pulse duration down to 3.8~ns, corresponding to a peak power of 1.6 MW with near-diffraction-limited beam quality. The compact and robust system was used for laser-induced breakdown spectroscopy (LIBS) experiments, demonstrating its practical usability. These results represent, to the best of our knowledge, the first demonstration of a Ho:YAG thin-disk laser providing MW peak-power in nanosecond regime.

\end{abstract*}

\section{Introduction}
Thin-disk lasers have become the workhorses of high-power and high-energy laser systems, alongside slab and fiber lasers \cite{giesen2007, fattahi2014, saraceno2019}. Their advantageous geometry, which enables efficient heat dissipation, allows power scaling to kilowatt level without compromising beam quality. Yb-doped materials, typically Yb:YAG, operating near 1~\textmu m have been successfully implemented in thin-disk architectures and have demonstrated tremendous progress over the past decades \cite{smrz2023}. In addition, recent advances in nonlinear pulse compression using multi-pass cells have made it possible to achieve ultrashort, femtosecond pulses \cite{viotti2022, pfaff2023}. This trend may eventually replace Ti:sapphire lasers, which have long been the standard for ultrashort pulse generation.
Consequently, Yb thin-disk lasers are commonly used as drivers for wavelength extension via nonlinear wavelength conversion techniques, such as harmonic generation for shorter wavelengths \cite{Dietrich2017, turcicova2022} and optical parametric oscillation/amplification (OPO/OPA) for longer wavelengths \cite{csanakova2024}. However, these methods suffer from limited conversion efficiency and require meticulous alignment. As Yb thin-disk lasers approach their performance limits, extending the thin-disk concept to other gain media that can directly generate other wavelengths has attracted growing interest \cite{tomilov2021}.
In particular, the short-wave infrared (SWIR) region around 2~\textmu m is of great interest due to its unique applications in spectroscopy \cite{thire2023}, remote sensing \cite{kuan2023}, and material processing \cite{Scholle10}. Scientific applications include driver sources for OPA \cite{malevich2016}, high-harmonic generation \cite{walke2025}, laser-Compton scattering \cite{hornberger2021, vlad2022}, and extreme-ultraviolet generation via laser-produced plasma \cite{mostafa2023}. Among various candidate materials \cite{zhang2018, zhang2025} , we have recently initiated studies on Ho:YAG thin-disk lasers. Here, we report our experimental results on a pulsed Ho:YAG thin-disk laser operated under Q-switching and cavity-dumping schemes.  

\section{Materials and Methods}

\begin{figure}[H]
\includegraphics[width=0.9\textwidth]{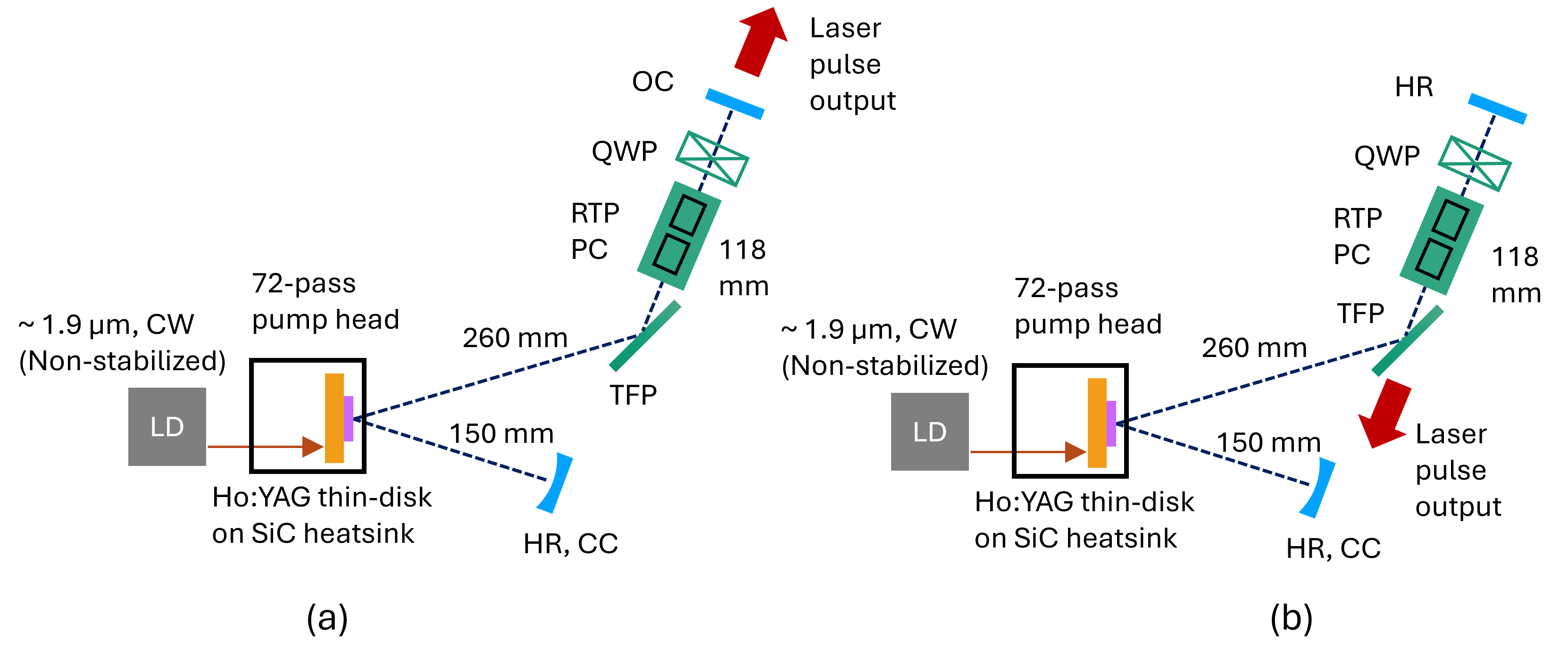}
\caption{Schematic of the experimental setup for (a) Q-switching and (b) cavity-dumping operations. OC: output coupler; QWP: quarter waveplate; RTP: rubidium titanyle phosphate; PC: Pockels cell; TFP: thin film polarizer; HR: high-reflective; CC: concave.\label{fig1}}
\end{figure}

A Ho:YAG thin-disk crystal with doping concentration of 1.5~at.\% and thickness of 400 ~\textmu m was used. The doping level is higher than that generally used in rod lasers \cite{qian2019} to compensate for the low gain inherent to the thin-disk geometry. However, higher doping also leads to undesirable effects such as increased heat generation, energy-transfer upconversion (ETU), and excited-state absorption (ESA) \cite{barnes2003, tomilov2024}. Regarding the thickness, thinner crystal would improve the heat dissipation but reduce both the gain and the pump absorption. Sufficient pump absorption is important, as residual pump light may be reflected back to the pump source. To ensure safe operation and higher gain, we selected a comparatively thick crystal in this study.
The heatsink was made of silicon carbide (SiC) which has three times lower thermal conductivity than chemical vapor deposited (CVD) diamond, but offers cost-effective alternative \cite{cvrcek2022}. The Ho:YAG thin-disk was mounted in a 72-pass pumping head. The pump source was a commercial LD manufactured by QPC Lasers Inc., delivering up to 50~W of continuous-wave power at around 1.91 ~\textmu m, corresponding to in-band pumping. However, it was not wavelength-stabilized with a volume Bragg grating (VBG) resulting in a broadband emission (12~nm full width at half maximum (FWHM)), and a temperature-dependent center wavelength. 
For Q-switching and cavity-dumping operations, a pair of rubidium titanyle phosphate (RTP) crystals with dimensions of 6×6×10~mm$^3$ was used as a Pockels cell (PC), driven by a high voltage (HV) driver with risetime below 5~ns. 
The experimental setup is depicted in Figure~\ref{fig1}. It is a simple setup with one arm containing a polarizer and an optical switch. 

In Q-switching, the pump energy is first stored in the gain medium. Ho:YAG possesses a long upper-state lifetime of about 7~ms \cite{sulc2021}, which is advantageous for Q-switching. When a HV is applied to the PC, the resonator becomes low-loss (i.e., high-Q), allowing laser buildup and pulse emission through the OC. OCs with measured transmittance of 2.6\%, 4.6\%, 5.8\%, and 7.1\% were tested, and the best performance was obtained with the 4.6\% OC. In cavity-dumping, the OC is replaced with an HR mirror. When the intracavity energy reaches its maximum, the PC is switched off, and the stored energy is dumped through the polarizer within one cavity round-trip.

\section{Results and Discussion}
\subsection{Q-switching}

\begin{figure}[H]
\includegraphics[width=1\textwidth]{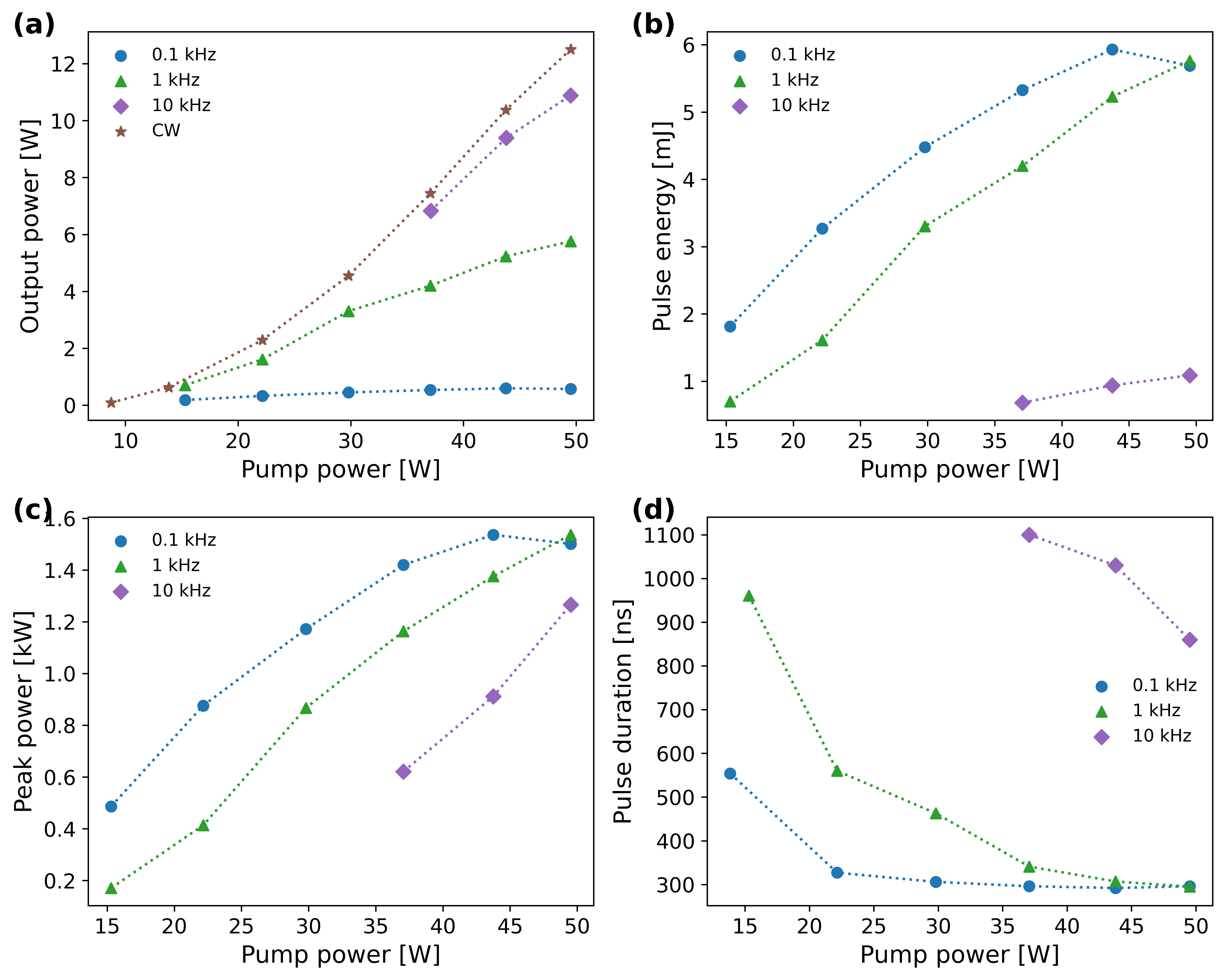}
\caption{(a) Average power, (b) pulse energy, (c) peak power, and (d) pulse duration at repetition rate of 0.1, 1, and 10~kHz. The CW power is also plotted on (a).\label{fig2}}
\end{figure}

The results of the Q-switched laser operation are shown in Figure~\ref{fig2}.
At a repetition rate of 10~kHz, the pulse energy was unstable at low pump power. 
Pulse energies exceeding 5~mJ were obtained at repetition rates of 1~kHz or lower. A clear rollover of the output pulse energy was observed near maximum pump level, which can be attributed to the non-stabilized pump spectrum, the elevated temperature (approximately 100~$^\circ$C according to thermal camera measurement) of the Ho:YAG thin-disk crystal and the RTP Pockels cell, and ETU process. The pulse duration decreased with increasing pump power due to the higher gain. The pulse duration remained longer than 292~ns, which limited the maximum peak power to 18.3~kW. To obtain shorter pulses, a higher net gain is required, which is inherently limited by the thin-disk geometry. Therefore, cavity-dumping is considered a more suitable approach for generating higher-peak-power pulses in thin-disk lasers.

\subsection{Cavity-dumping}

\begin{figure}[H]
\includegraphics[width=1\textwidth]{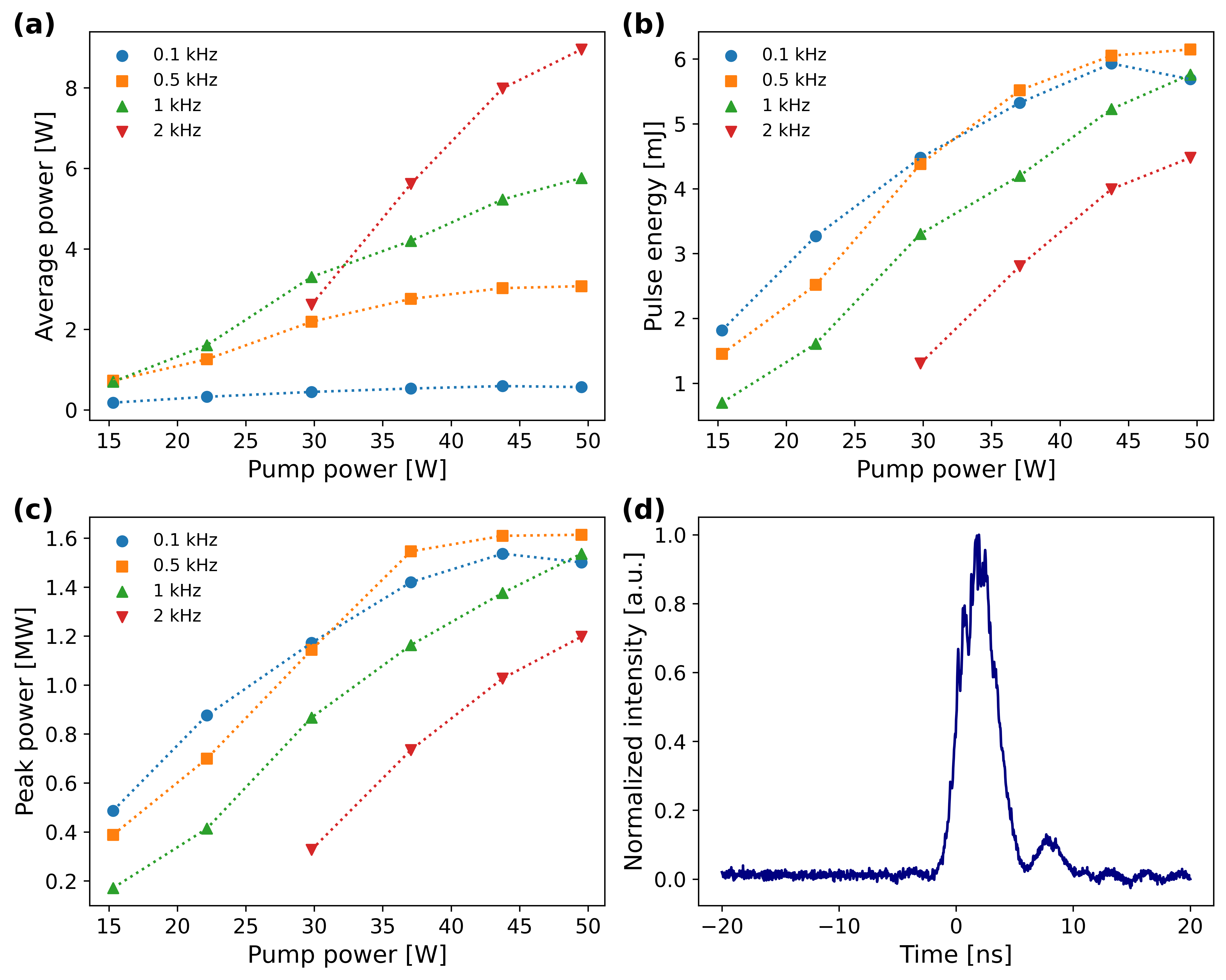}
\caption{(a) Average power, (b) pulse energy, and (c) peak power at different repetition rates. (d) Temporal waveform of a cavity-dumped pulse.\label{fig3}}
\end{figure}

\begin{figure}[H]
\includegraphics[width=0.9\textwidth]{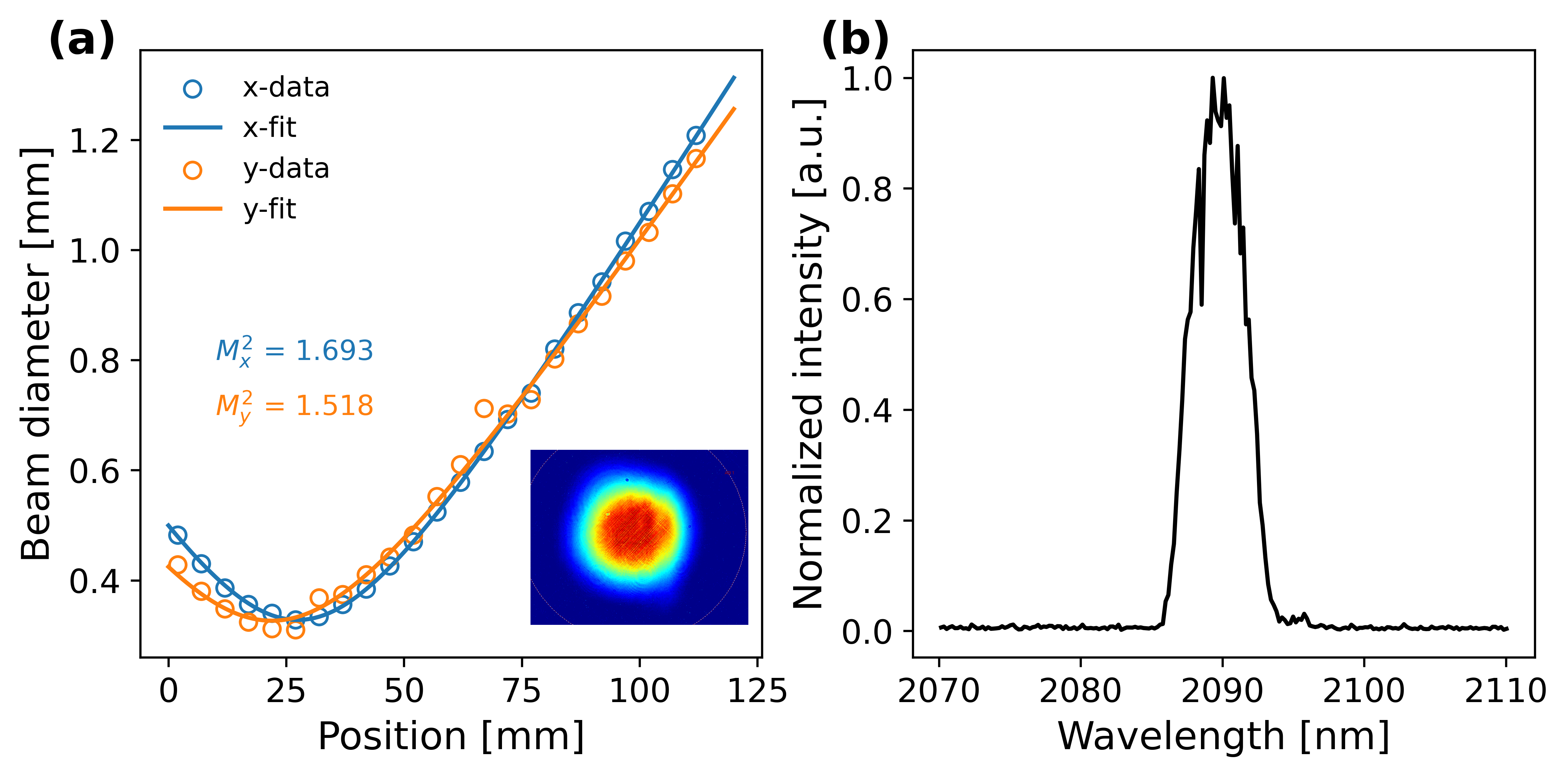}
\caption{(a) M$^2$ measurement conforming ISO 11146-1. The inset shows the beam profile after the beam expander. (b) Spectrum of the cavity-dumped pulse.\label{fig4}}
\end{figure}

The results of the cavity-dumped operation at different repetition rates, including the average power, pulse energy, peak power, and a representative temporal waveform, are shown in Figure~\ref{fig3}.
The pulse duration was approximately 3.8~ns in all cases, corresponding to the round-trip time of the 114~cm optical cavity. The two-orders-of-magnitude shorter pulse duration compared with the Q-switched pulses enabled the generation of peak powers up to 1.6~MW. The beam quality, characterized by the M$^2$ factor, was evaluated by measuring the beam caustic using a 150~mm focal-length lens and a beam profiler. The measured M$^2$ values were approximately 1.7 in the horizontal and 1.5 in the vertical directions at a repetition rate of 0.5~kHz and a peak power of 1.5~MW, as shown in Figure~\ref{fig4}, together with the emission spectrum measured using a rotating grating spectrometer (APE, waveScan). The inset of Figure~\ref{fig4}(a) shows the beam profile after the beam expander. The beam was expanded to achieve higher intensity upon focusing for LIBS experiments \cite{vozar2026}. 

\subsection{Energy scaling}

In order to further increase the pulse energy, four strategies can be considered. Firstly, the simplest approach is to use a Ho:YAG thin-disk bonded to a CVD-diamond heat sink instead of SiC. The threefold higher thermal conductivity of diamond would help reduce the crystal temperature. In Ho:YAG, maintaining a low temperature is even more crucial than in Yb:YAG, since elevated temperature enhances ETU. 
Secondly, a straightforward improvement is to employ a narrow-linewidth, wavelength-stabilized pump source. However, high-power LDs at the 1.9~\textmu m wavelength remain niche in the market; to our knowledge, 60 W is currently the highest power commercially available. Therefore, an alternative and more practical approach is to develop a Tm-fiber laser tuned to 1907~nm, which can be pumped by 793~nm LDs. This configuration would enhance pump absorption and reduce thermal loading, leading to more efficient laser operation. 
Thirdly, the number of passes through the thin-disk gain medium per round-trip can be increased \cite{radmard2023}. For instance, introducing two reflections instead of one effectively doubles the number of pass (from 4 to 8) per round-trip. While this is expected to increase the pulse energy and shorten the Q-switched pulse duration, in cavity-dumping operation the longer cavity length would result in longer pulse duration. Finally, optimization of the doping concentration and crystal thickness is essential. A thinner crystal provides better heat dissipation at the cost of lower gain and reduced pump absorption. Conversely, a higher doping concentration increases the gain but also leads to stronger heating and ETU. Therefore, achieving an appropriate balance between doping concentration and thickness is critical for further energy scaling.

\section{Conclusions}

We have demonstrated Q-switching and cavity-dumping operation of a Ho:YAG thin-disk laser pumped by a 1.9~\textmu m LD. The laser delivered more than 5~mJ of pulse energy in Q-switching at repetition rates of 1~kHz or lower. The shortest pulse duration achieved was 292~ns, and the highest peak power was 18.3~kW. By employing cavity-dumping, the pulse duration was significantly reduced to 3.8~ns, which is determined by the cavity round-trip time, leading to a peak power as high as 1.6~MW. These results confirm that cavity-dumping Ho:YAG thin-disk laser is an effective approach for generating high-peak-power nanosecond pulses around 2.1~\textmu m. Further performance improvements can be expected by using a CVD-diamond heat sink, a wavelength-stabilized or Tm-fiber-based pump source, multi-pass geometry, and optimized doping concentration and disk thickness. In addition, the laser was used for LIBS experiments \cite{vozar2026}. Encouraged by these results, we plan to leverage the Ho:YAG thin-disk as a main amplifier in a chirped-pulse amplification (CPA) scheme, aiming for picosecond pulses with millijoule-level pulse energies.

\begin{backmatter}
\bmsection{Funding}
This work was supported by MERIT programme 101081195 powered by Central Bohemian Innovation Center and co-funded by the European Union and state budget of the Czech Republic under project LasApp CZ.02.01.01/00/22\_008/0004573.
This research was co-funded by the European Union (MERIT - Grant Agreement No. 101081195) and the state budget of the Czech republic under the project LasApp CZ.02.01.01/00/22\_008/0004573.


\bmsection{Disclosures}
The authors declare no conflicts of interest.

\bmsection{Data availability} Data underlying the results presented in this paper may be obtained from the authors upon reasonable request.


\end{backmatter}




\bibliography{sample}

\begin{thebibliography}{10}
\newcommand{\enquote}[1]{``#1''}

\bibitem{giesen2007}
A.~Giesen and J.~Speiser, \enquote{Fifteen years of work on thin-disk lasers: Results and scaling laws,} {\protect\JournalTitle{IEEE Journal on Selected Topics in Quantum Electronics}} \textbf{13}, 598--609 (2007).

\bibitem{fattahi2014}
H.~Fattahi, H.~G. Barros, M.~Gorjan, \emph{et~al.}, \enquote{Third-generation femtosecond technology,} {\protect\JournalTitle{Optica}} \textbf{1}, 45--63 (2014).

\bibitem{saraceno2019}
C.~J. Saraceno, D.~Sutter, T.~Metzger \emph{et~al.}, \enquote{The amazing progress of high-power ultrafast thin-disk lasers,} {\protect\JournalTitle{J. Eur. Opt. Soc.-Rapid Publ.}} \textbf{15}, 15 (2019).

\bibitem{smrz2023}
M.~Smrž, J.~Mužik, and S.~Nagisetty, \enquote{Yag lasers for lithography and metrology,} in \emph{Photon Sources for Lithography and Metrology,}  V.~Bakshi, ed. (SPIE Press, Bellingham, WA, USA, 2023), chap.~22, pp. 911--997.

\bibitem{viotti2022}
A.-L. Viotti, M.~Seidel, E.~Escoto, \emph{et~al.}, \enquote{Multi-pass cells for post-compression of ultrashort laser pulses,} {\protect\JournalTitle{Optica}} \textbf{9}, 197--216 (2022).

\bibitem{pfaff2023}
Y.~Pfaff, G.~Barbiero, M.~Rampp, \emph{et~al.}, \enquote{Nonlinear pulse compression of a 200 mj and 1 kw ultrafast thin-disk amplifier,} {\protect\JournalTitle{Optics Express}} \textbf{31}, 22740--22756 (2023).

\bibitem{Dietrich2017}
T.~Dietrich, S.~Piehler, M.~Rumpel, \emph{et~al.}, \enquote{{Highly-efficient continuous-wave intra-cavity frequency-doubled Yb:LuAG thin-disk laser with 1 kW of output power},} {\protect\JournalTitle{Opt. Express}} \textbf{25}, 4917--4925 (2017).

\bibitem{turcicova2022}
H.~Turcicova, O.~Novak, J.~Muzik, \emph{et~al.}, \enquote{Laser induced damage threshold (lidt) of $\beta$-barium borate (bbo) and cesium lithium borate (clbo)--overview,} {\protect\JournalTitle{Optics \& Laser Technology}} \textbf{149}, 107876 (2022).

\bibitem{csanakova2024}
B.~Csanakov{\'a}, O.~Nov{\'a}k, L.~Ro{\v{s}}kot, \emph{et~al.}, \enquote{High power single crystal kta optical parametric amplifier for efficient 1.4--3.5 $\mu$m mid-ir radiation generation,} {\protect\JournalTitle{Laser Physics}} \textbf{34}, 075401 (2024).

\bibitem{tomilov2021}
S.~Tomilov, M.~Hoffmann, Y.~Wang, and C.~J. Saraceno, \enquote{Moving towards high-power thin-disk lasers in the 2 $\mu$m wavelength range,} {\protect\JournalTitle{J. Phys. Photonics}} \textbf{3}, 022002 (2021).

\bibitem{thire2023}
N.~Thiré, G.~Chatterjee, Y.~Pertot, \emph{et~al.}, \enquote{A versatile high-average-power ultrafast infrared driver tailored for high-harmonic generation and vibrational spectroscopy,} {\protect\JournalTitle{Sci. Rep.}} \textbf{13}, 18874 (2023).

\bibitem{kuan2023}
K.~Li, C.~Niu, C.~Wu, \emph{et~al.}, \enquote{Development of a 2~\textmu m solid-state laser for lidar in the past decade,} {\protect\JournalTitle{Sensors}} \textbf{23} (2023).

\bibitem{Scholle10}
K.~Scholle, S.~Lamrini, P.~Koopmann, and P.~Fuhrberg, \enquote{2 \textmu m laser sources and their possible applications,} in \emph{Frontiers in Guided Wave Optics and Optoelectronics,}  B.~Pal, ed. (IntechOpen, London, 2010), chap.~21.

\bibitem{malevich2016}
P.~Malevich, T.~Kanai, H.~Hoogland, \emph{et~al.}, \enquote{Broadband mid-infrared pulses from potassium titanyl arsenate/zinc germanium phosphate optical parametric amplifier pumped by tm, ho-fiber-seeded ho:yag chirped-pulse amplifier,} {\protect\JournalTitle{Opt. Lett.}} \textbf{41}, 930--933 (2016).

\bibitem{walke2025}
D.~Walke, A.~Ko\c{c}, F.~Gores, \emph{et~al.}, \enquote{High-average-power, few-cycle, 2.1 {\textmu}m opcpa laser driver for soft-x-ray high-harmonic generation,} {\protect\JournalTitle{Opt. Express}} \textbf{33}, 10006--10019 (2025).

\bibitem{hornberger2021}
B.~Hornberger, J.~Kasahara, R.~Ruth, \emph{et~al.}, \enquote{{Inverse Compton scattering X-ray source for research, industry and medical applications},} in \emph{International Conference on X-Ray Lasers 2020,}  vol. 11886 D.~Bleiner, ed., International Society for Optics and Photonics (SPIE, 2021), p. 1188609.

\bibitem{vlad2022}
V.~Mușat, A.~Latina, and G.~D’Auria, \enquote{A high-energy and high-intensity inverse compton scattering source based on compactlight technology,} {\protect\JournalTitle{Photonics}} \textbf{9} (2022).

\bibitem{mostafa2023}
Y.~Mostafa, L.~Behnke, D.~J. Engels, \emph{et~al.}, \enquote{Production of 13.5 nm light with 5\% conversion efficiency from 2 \textmu m laser-driven tin microdroplet plasma,} {\protect\JournalTitle{Appl. Phys. Lett.}} \textbf{123}, 234101 (2023).

\bibitem{zhang2018}
J.~Zhang, F.~Schulze, K.~F. Mak, \emph{et~al.}, \enquote{High-power, high-efficiency tm:yag and ho:yag thin-disk lasers,} {\protect\JournalTitle{Laser \& Photonics Reviews}} \textbf{12}, 1700273 (2018).

\bibitem{zhang2025}
L.~Zhang, B.~Lei, Y.~Cheng, \emph{et~al.}, \enquote{Hundred-watt-level cryogenic near diffraction-limited composite thin-disk ho:ylf oscillator at 2 µm,} {\protect\JournalTitle{Opt. Express}} \textbf{33}, 20699--20708 (2025).

\bibitem{qian2019}
C.~Qian, B.~Yao, Y.~Ju, \emph{et~al.}, \enquote{110.4 mj, 1 khz repetition rate, ho:yag master oscillator power amplifier,} {\protect\JournalTitle{Appl. Opt.}} \textbf{58}, 879--882 (2019).

\bibitem{barnes2003}
N.~P. Barnes, B.~M. Walsh, and E.~D. Filer, \enquote{Ho: Ho upconversion: applications to ho lasers,} {\protect\JournalTitle{J. Opt. Soc. Am. B}} \textbf{20}, 1212--1219 (2003).

\bibitem{tomilov2024}
S.~Tomilov, P.~Loiko, K.~Eremeev, \emph{et~al.}, \enquote{Modeling and optimization of in-band pumped ho:yag lasers for high-power operation,} in \emph{Laser Congress 2024 (ASSL, LAC, LS\&C), Technical Digest Series,}  (Optica Publishing Group, 2024), p. JTu2A.6.

\bibitem{cvrcek2022}
J.~Cvr{\v{c}}ek, M.~Cimrman, D.~Vojna, \emph{et~al.}, \enquote{Investigation of the lasing performance of a crystalline-coated yb:yag thin-disk directly bonded onto a silicon carbide heatsink,} {\protect\JournalTitle{Opt. Express}} \textbf{30}, 7708--7715 (2022).

\bibitem{sulc2021}
J.~Šulc, M.~Němec, D.~Vyhl{\'i}dal, \emph{et~al.}, \enquote{{Holmium doping concentration influence on Ho:YAG crystal spectroscopic properties},} in \emph{Solid State Lasers XXX: Technology and Devices,}  vol. 11664 W.~A. Clarkson and R.~K. Shori, eds., International Society for Optics and Photonics (SPIE, 2021), p. 1166413.

\bibitem{vozar2026}
T.~Vozár, L.~Čechová, J.~Buday, \emph{et~al.}, \enquote{Comparison of two laser wavelengths for libs bioimaging of plants grown in lunar regolith,} {\protect\JournalTitle{Spectrochim. Acta, Part B}}  (2025). Submitted.

\bibitem{radmard2023}
S.~Radmard, A.~Moshaii, K.~Pasandideh, and S.~Arabgari, \enquote{Introducing a 2v-resonator for the improvement of pulse stability in a high-gain q-switched yb: Yag thin-disk laser,} {\protect\JournalTitle{Optics Express}} \textbf{31}, 12128--12137 (2023).

\end{thebibliography}






\end{document}